\documentclass[
superscriptaddress,
amsmath,amssymb, aps, prl, 
reprint, 
floatfix,
]{revtex4-2}

\usepackage{graphicx}
\usepackage{dcolumn}
\usepackage{bm}
\usepackage{hyperref}
\usepackage{siunitx}
\usepackage{subcaption}

\DeclareSIUnit\sr{sr}

\newcommand{\gamfive}{\ensuremath{\gamma = 10^{5}}}
\newcommand{\fluxfive}{\ensuremath{\num{2.6e-16}\,\unit{\cm^{-2}.\s^{-1}.\sr^{-1}}}}
\newcommand{\LimitFiveExp}{1378.7\,^{+589.7}_{-404.1}}
\newcommand{\LimitFiveObs}{2557.0} 

\newcommand{\LimitSixExp}{25.3\,^{+15.1}_{-8.9}}
\newcommand{\LimitSixObs}{22.4}

\newcommand{\gamseven}{\ensuremath{\gamma = 10^{7}}}
\newcommand{\fluxseven}{\ensuremath{\num{8.8e-19}\,\unit{\cm^{-2}.\s^{-1}.\sr^{-1}}}}
\newcommand{\LimitSevenExp}{8.8\,^{+1.7}_{-0.2}}
\newcommand{\LimitSevenObs}{8.8}

\begin{document}
\preprint{APS/123-QED}

\title{Search for Ultra-Relativistic Magnetic Monopoles with LHAASO-KM2A}

\author{Zhen Cao}
\affiliation{State Key Laboratory of Particle Astrophysics \& Experimental Physics Division \& Computing Center, Institute of High Energy Physics, Chinese Academy of Sciences, 100049 Beijing, China}
\affiliation{University of Chinese Academy of Sciences, 100049 Beijing, China}
\affiliation{Tianfu Cosmic Ray Research Center, 610000 Chengdu, Sichuan,  China}
 
\author{F. Aharonian}
\affiliation{Tianfu Cosmic Ray Research Center, 610000 Chengdu, Sichuan,  China}
\affiliation{University of Science and Technology of China, 230026 Hefei, Anhui, China}
\affiliation{Yerevan State University, 1 Alek Manukyan Street, Yerevan 0025, Armenia }
\affiliation{Max-Planck-Institut for Nuclear Physics, P.O. Box 103980, 69029  Heidelberg, Germany}
 
\author{Y.X. Bai}
\affiliation{State Key Laboratory of Particle Astrophysics \& Experimental Physics Division \& Computing Center, Institute of High Energy Physics, Chinese Academy of Sciences, 100049 Beijing, China}
\affiliation{Tianfu Cosmic Ray Research Center, 610000 Chengdu, Sichuan,  China}
 
\author{Y.W. Bao}
\affiliation{Tsung-Dao Lee Institute \& School of Physics and Astronomy, Shanghai Jiao Tong University, 200240 Shanghai, China}
 
\author{D. Bastieri}
\affiliation{Center for Astrophysics, Guangzhou University, 510006 Guangzhou, Guangdong, China}
 
\author{X.J. Bi}
\affiliation{State Key Laboratory of Particle Astrophysics \& Experimental Physics Division \& Computing Center, Institute of High Energy Physics, Chinese Academy of Sciences, 100049 Beijing, China}
\affiliation{University of Chinese Academy of Sciences, 100049 Beijing, China}
\affiliation{Tianfu Cosmic Ray Research Center, 610000 Chengdu, Sichuan,  China}
 
\author{Y.J. Bi}
\affiliation{State Key Laboratory of Particle Astrophysics \& Experimental Physics Division \& Computing Center, Institute of High Energy Physics, Chinese Academy of Sciences, 100049 Beijing, China}
\affiliation{Tianfu Cosmic Ray Research Center, 610000 Chengdu, Sichuan,  China}
 
\author{W. Bian}
\affiliation{Tsung-Dao Lee Institute \& School of Physics and Astronomy, Shanghai Jiao Tong University, 200240 Shanghai, China}
 
\author{J. Blunier}
\affiliation{APC, Universit'e Paris Cit'e, CNRS/IN2P3, CEA/IRFU, Observatoire de Paris, 119 75205 Paris, France}
 
\author{A.V. Butkevich}
\affiliation{Institute for Nuclear Research of Russian Academy of Sciences, 117312 Moscow, Russia}
 
% ...
\collaboration{The LHAASO Collaboration}
 
\date{\today}

\begin{abstract}
Ultra-relativistic magnetic monopoles traversing the atmosphere are expected to initiate extensive air showers characterized 
by a muon deficit and a compact lateral profile. Using 1438.81 live days of data from the LHAASO-KM2A array, 
we present a search for these cosmic particles within the Lorentz factor range of $10^5$ to $10^7$. 
A machine-learning classifier utilizing lateral shower shape observables from the electromagnetic detectors is deployed to 
separate the signal from the hadronic background. No statistical excess above the background expectation is observed, 
and upper limits at the 90\% confidence level are established on the isotropic monopole flux. 
The resulting upper limit is \fluxfive{} at \gamfive{} and reaches \fluxseven{} at \gamseven{} 
which improves upon the astrophysical Parker bound by over three orders of magnitude. 
This work establishes constraints on the monopole flux within a previously unconstrained velocity window.
\end{abstract}

\makeatletter
\newcommand{\orderedcorrespondence}[1]{%
  \begingroup\sanitize@url\@makeother\@\orderedcorrespondence@{#1}%
}
\def\orderedcorrespondence@#1#2{%
  \endgroup
  \g@addto@macro\@FMN@list{%
    \@TBN@opr{#1}{\@join{\@separator}{}{\href{mailto:#2}{#2}}}}%
}

\orderedcorrespondence{1}{zh_zhang@mail.ustc.edu.cn}
\orderedcorrespondence{2}{wuyusheng@ustc.edu.cn}
\orderedcorrespondence{3}{yangrz@ustc.edu.cn}
\orderedcorrespondence{4}{yuyanhong@htu.edu.cn}
\makeatother
\maketitle


\textit{Introduction.}---Magnetic monopoles (MMs) are hypothetical particles originally postulated 
by Dirac~\cite{Dirac1931} in 1931 to restore electromagnetic symmetry and explain the quantization of electric charge. 
They were subsequently shown to be mandatory topological defects arising from spontaneous symmetry breaking 
in Grand Unified Theories~\cite{tHooft1974, Polyakov1974}. The discovery of MMs would have important implications for fundamental physics, 
shedding light on charge quantization mechanisms, phase transitions in the early Universe—such as cosmic inflation and 
the associated cosmological monopole problem~\cite{Preskill1979, Guth1981}—and the ultimate unification of forces at inaccessible energy scales. 
While MMs are generally expected to be massive and stable, their exact phenomenology remains subject to substantial theoretical uncertainty due 
to their strong-coupling nature and unknown mass scale. Experimentally,
MMs carrying a single unit of magnetic charge interact with ordinary matter with an electromagnetic coupling strength significantly 
larger than that of an electron. This key feature provides a clear experimental signature through 
distinctive energy-loss mechanisms~\cite{Ahlen1978}, which have been widely utilized to search for MMs.

Over the past few decades, numerous experiments have been conducted to search for MMs, 
primarily categorized into accelerator-based and cosmic-ray-based experiments. 
Collider searches are sensitive to MMs with masses ranging from the GeV to TeV scales, probing production cross sections down to the 
femtobarn level~\cite{ATLAS:2023ofz, MoEDAL:2021vix}. 
Conversely, cosmic-ray searches can probe much larger mass scales and are sensitive 
to extremely rare fluxes down to the order of $\num{1.0e-21}\,\unit{\cm^{-2}.\s^{-1}.\sr^{-1}}$~\cite{Auger2016}. 
Grand-unified-theory (GUT) monopoles are generally expected at very high mass scales, 
while intermediate-mass monopoles (IMMs) may arise from symmetry breaking below the grand-unification scale,
with model-dependent masses set by the corresponding symmetry-breaking scales~\cite{Kephart2017}.
Their relatively lower masses allow IMMs to be accelerated to relativistic or ultra-relativistic velocities by cosmic magnetic fields~\cite{Auger2016}.
There exists also the astrophysical Parker bound~\cite{Parker1970}, based on the survival of the Milky Way's magnetic field, 
which constrains the monopole flux to the order of \(\num{1.0e-15}\,\unit{\cm^{-2}.\s^{-1}.\sr^{-1}}\)~\cite{Turner1982, Adams:1993fi}.

Currently, the leading limits on the cosmic monopole flux are established by different detectors depending on the velocity regime. 
The MACRO experiment~\cite{MACRO2002} sets the most competitive constraints over a wide non-relativistic regime, 
and extends into the relativistic region alongside IceCube~\cite{IceCube2016} for velocities up to $\beta \approx 0.995$. 
In the ultra-relativistic domain, the RICE~\cite{RICE2006} and the Pierre Auger Observatory~\cite{Auger2016} dominate at extreme energy scales with Lorentz factors $\gamma \geq 10^{7}$. 
However, a gap remains at lower Lorentz factors, $10^5 \leq \gamma \leq 10^7$,
where the detection mechanisms employed by these cosmic-ray observatories limit their sensitivity,
leaving this region without competitive constraints on the monopole flux.

In this Letter, we present the first search for such cosmic MMs using 
the Large High Altitude Air Shower Observatory square-kilometer array (LHAASO-KM2A)~\cite{He2018}. With its large fiducial volume 
and continuous exposure, the array is well-positioned to fill this phase-space gap. 
Its dense, co-located arrays of electromagnetic detectors (EDs) and muon detectors (MDs) 
provide strong background rejection and fine spatial granularity, making it well-suited to search for MMs with Lorentz factors between $10^{5}$ 
and $10^{7}$.
In this ultra-relativistic window, penetrating MMs are expected to initiate extensive air showers through continuous interactions. 
For this study, a monopole mass of $10^5\,\mathrm{GeV}$ is chosen as a reference benchmark following the energy-loss study of Wick et al.~\cite{Wick2003}.
According to the energy loss formulas of monopoles~\cite{Ahlen1978, Wick2003}, 
bremsstrahlung radiation scales inversely with the mass at fixed $\gamma$ and is heavily suppressed compared to pair production 
and photonuclear effects at this scale. 
The latter two dominant processes depend primarily on \(\gamma\), making the shower development approximately mass independent.
Consequently, the results of this study can be approximately interpreted for higher monopole masses provided that 
bremsstrahlung and the change in the monopole Lorentz factor during atmospheric propagation remain negligible.


\textit{Detector and Simulation.}---%
The LHAASO observatory, situated at an altitude of $4410\,\text{m}$ above sea level, 
is dedicated to high-energy astronomy, designed for gamma-ray observations from $10\,\text{TeV}$ to $10\,\text{PeV}$ 
and cosmic-ray measurements up to $100\,\text{PeV}$~\cite{He2018}. This analysis utilizes data recorded by KM2A, 
which comprises 5216 EDs and 1188 underground MDs. The dense layout of the ED array provides high-granularity sampling of the electromagnetic 
component of air showers, enabling precise reconstruction of both the core position and the lateral distribution profile~\cite{LHAASO:2021crab}.
Concurrently, the extensive coverage of the MD array allows for an efficient measurement of the secondary muon content for each shower. 
The combination of these two subsystems offers powerful discrimination to distinguish hypothetical ultra-relativistic 
MMs—which are characterized by muon deficit and a compact lateral profile—from the hadronic cosmic-ray background. 
The dataset analyzed in this Letter was accumulated between July 2021 and July 2025 for a total of 1438.81 live days, 
corresponding to a stable operation period with the fully deployed array configuration~\cite{cao2024data}.

Monte Carlo (MC) simulations are performed using CORSIKA~\cite{Heck1998} integrated with CONEX~\cite{Bergmann2007conex, Pierog2008hybrid} 
to model the extensive air showers. 
Standard hadronic cosmic-ray backgrounds are generated using QGSJET-II-04~\cite{Ostapchenko2011} 
for high-energy regimes and GHEISHA~\cite{Fesefeldt1985} for low-energy components. 
For ultra-relativistic MMs with $\mathrm{M} = 10^5\text{ GeV}$, the continuous ionization energy loss is supplemented 
by stochastic processes~\cite{Ahlen1978,Wick2003}. Importantly, the total energy loss in this high-energy regime is dominated 
by pair production and photonuclear interactions, while bremsstrahlung radiation is suppressed by the large mass of monopole~\cite{Wick2003}. 
Signal samples are simulated at three discrete Lorentz factors: $\gamma = 10^5$, $10^6$, and $10^7$. 
Secondary particles reaching the ground are processed through G4KM2A~\cite{LHAASO2024g4km2a}, 
the official LHAASO detector simulation based on GEANT4~\cite{Agostinelli:2003geant4}, 
which accurately accounts for the realistic geometry, material budgets, and electronic trigger responses of the KM2A array.

\textit{Analysis Methodology.}---%
Extensive air showers induced by ultra-relativistic MMs develop primarily through electromagnetic pathways. 
The absence of strong hadronic interactions will result in a deficit of secondary muons at the LHAASO observation 
level compared to standard cosmic-ray events. Furthermore, owing to their stable characteristics, monopoles continuously 
deposit energy through sequential interactions with atmospheric nuclei along their trajectories~\cite{Auger2016}. 
This  energy drain induces multiple overlapping sub-showers, which reconstruct as a single, highly compact shower profile. 
Crucially, unlike standard cosmic-ray cascades that undergo rapid atmospheric attenuation after reaching their developmental maxima, 
monopole showers exhibit a persistent growth tendency sustained by these continuous secondary injections~\cite{Wick2003}. 
Consequently, monopole-induced cascades exhibit a younger shower age ($s_{\mathrm{age}}$) and a denser lateral distribution. 
This distinctive penetrating depth implies that for large zenith angles ($\theta_{\mathrm{rec}}$), 
where backgrounds are heavily attenuated by the extended atmospheric grammage, 
the survival and detection probabilities of isotropic MMs remain extremely high, 
providing a topological signature for experimental discrimination.

Owing to the dominance of electromagnetic energy loss, 
monopole-induced showers behave similarly to those initiated by high-energy gamma rays and
cosmic-ray electrons--positrons ($e^\pm$). These electromagnetic primaries therefore
constitute potential backgrounds to the monopole search. To suppress the
dominant localized gamma-ray emission while preserving signal acceptance, the
search excludes the Galactic plane within $|b|<10^\circ$ and masks all
cataloged Galactic and extragalactic gamma-ray sources
~\cite{LHAASO:2024catalog}. This spatial masking benefits from the excellent
angular resolution of KM2A, which is better than $0.2^\circ$ above
$100~\mathrm{TeV}$ as determined from Crab Nebula observations~\cite{LHAASO:2024performance}.
As the $e^\pm$ spectrum has not been measured directly at the energies relevant to this search, 
we estimate this background by extrapolating the all-electron spectrum measured by H.E.S.S.~\cite{HESS:2024CRE}. 
Under this extrapolation, the expected $e^\pm$ contribution within the SR is approximately $10^{-3}$ of the hadronic cosmic-ray background 
and is therefore neglected. H.E.S.S. finds both the diffuse extragalactic and high-latitude Galactic gamma-ray fluxes 
below the $e^\pm$ flux in the energy range of its measurement~\cite{HESS:2024CRE}, 
while the isotropic diffuse gamma-ray spectrum measured by Fermi-LAT falls steeply above a few hundred GeV~\cite{Ackermann2015IGRB}.
Accordingly, the residual diffuse gamma-ray background is also neglected.

To separate the rare magnetic monopole signals from the overwhelming hadronic cosmic-ray background, 
basic shower quality selections are applied, and the signal region (SR) is established using two complementary observables: 
the reconstructed shower energy ($E_{\mathrm{rec}}$) and the muon-to-electromagnetic ratio ($R_{\mu e}$). 
The former, $E_{\mathrm{rec}}$, is reconstructed via an XGBoost model~\cite{Chen:2016xgboost} trained specifically on 
simulated monopole samples. Utilizing a linear-fit baseline, 
the model takes the reconstructed zenith angle and the lateral particle density at $50~\text{m}$ from the shower core ($\rho_{50}$) as inputs, 
where $\rho_{50}$ is obtained from an NKG-function fit~\cite{Kamata1958,Greisen:1960progress} to the shower front. 
These two observables establish a log-linear scaling between the shower size and the reconstructed energy, 
aligning the energy scales of the background and signal events to enhance the signal purity for specific kinematic configurations. 
For $\gamma = 10^5$, the corresponding MC proton background energy centers around 160~TeV and increases by one order of magnitude for 
each subsequent $\gamma$ point. 
The \(10^8~\mathrm{GeV}\) normalization in Table~\ref{tab:cuts} is chosen for scaling monopole energy to have a numerical magnitude comparable to that of the corresponding cosmic-ray background on the logarithmic scale.
The latter, $R_{\mu e}$, parameterizes the secondary muon content and is defined as $R_{\mu e} = \log_{10}[(N_\mu + 10^{-4})/N_{\mathrm{EM}}]$, 
where $N_\mu$ and $N_{\mathrm{EM}}$ denote the particle counts within $15
\sim 200~\text{m}$ for MDs and $400~\text{m}$ for EDs  from the shower core~\cite{LHAASO:2021crab}, 
respectively. Because standard hadronic showers possess a higher muon yield, 
this ratio provides a useful handle to suppress the background. To maximize the experimental sensitivity, 
the specific selection thresholds for both observables are optimized independently for each discrete Lorentz factor $\gamma = 10^5$, $10^6$, 
and $10^7$, defining the customized SR boundaries and corresponding MMs survival fractions ($\varepsilon_{\mathrm{MMs}}$) summarized in 
Table~\ref{tab:cuts}.

\begin{table}[htbp]
    \caption{\label{tab:cuts}Optimized selection ranges for $\log_{10}(E_{\mathrm{rec}}/ [10^8 \,\text{GeV}])$ and $R_{\mu e}$, defining the SR and $\varepsilon_{\mathrm{MMs}}$ for each discrete Lorentz factor $\gamma$.}
    \begin{ruledtabular}
    \begin{tabular}{cccc}
    $\gamma$ & $\log_{10}(E_{\mathrm{rec}}/ [10^8 \,\text{GeV}])$ & $R_{\mu e}$ &  $\varepsilon_{\mathrm{MMs}}$\\
    \hline
    $10^5$ & $[1.95, 2.10]$ & $[-5.0, -2.9]$ & $16.65\%$ \\
    $10^6$ & $[2.85, 3.10]$ & $[-5.0, -2.6]$ & $71.37\%$ \\
    $10^7$ & $[3.75, 4.20]$ & $[-5.0, -2.2]$ & $97.10\%$ \\
    \end{tabular}
    \end{ruledtabular}
\end{table}

Events in the zero-muon limit ($R_{\mu e}<-5$) are excluded from the SR. For $\gamma = 10^5$, 
this is due to the much worsened signal-over-background ratio and 
the incompatibility to apply the data-driven background modeling templates extracted from the muon-populated control regions. 
For the higher Lorentz factors ($\gamma \ge 10^6$), the enormous energy scale of the showers makes zero-muon fluctuations negligible. 
Therefore, the exclusion of zero-muon events has no impact on the high-energy search 
while ensuring a consistent treatment across all investigated kinematic points.

The distributions of the primary selection observables are presented in Fig.~\ref{fig:sr_variables}. 
As illustrated, the $R_{\mu e}$ thresholds define an extreme muon-poor window, 
suppressing the background survival fraction to $5.5 \times 10^{-6}$ for \gamfive and lower for higher $\gamma$ after the $E_{rec}$ cut. 
Based on simulation, the residual cosmic-ray background surviving within this region is heavily dominated by primary protons weighted 
by the Gaisser composition model~\cite{Gaisser:2012h3a}, which account for over 99\% of the remaining hadronic component. 
However, due to this massive rejection power, the surviving MC sample size inside the SR is depleted to near-zero events, 
making direct simulation-driven background modeling highly unstable and infeasible. Furthermore, 
in this muon-poor regime near the SR boundary, predictions from alternative hadronic interaction models exhibit substantial discrepancies, 
limiting their reliability as a simulation baseline~\cite{Dembinski:2019muon, LHAASO2024g4km2a}. 

To circumvent these limitations, a novel data-driven approach~\cite{ATLAS:2020MultijetTemplate} is developed to evaluate the hadronic background,
removing dependencies on both the hadronic generators and primary mass composition assumptions. 
The method is grounded in an intrinsic feature of hadronic air showers within the control region (CR): 
at a fixed $E_{\mathrm{rec}}$ range, the shower lateral profiles stabilize asymptotically with decreasing $R_{\mu e}$. 
This asymptotic behavior is verified across six shape observables in the low-$R_{\mu e}$ region (see definition and Fig.~\ref{fig:asymptotic_scan}-\ref{fig:shape_stability} in End Matter). 
Based on this stability, the background shape inside the SR is parameterized using an empirical template extracted 
from a background-dominated data sample adjacent to the SR boundary, with $-2.9<R_{\mu e}<-2.85$ for the $\gamma=10^5$ search taken as an example.

To further exploit the topological differences between signal and background showers, 
a Multi-Layer Perceptron (MLP)~\cite{Pedregosa:2011scikit} is utilized for multivariate classification. 
The network takes six observables in Fig.~\ref{fig:shape_stability} as inputs to differentiate the simulated monopole signals 
from the data-driven background template. Training is performed using a 5-fold cross-validation strategy to ensure stability 
and mitigate overfitting, yielding a mean validation AUC of $0.968$. The output score of the trained MLP is subsequently employed as 
the final discriminant in the likelihood fit to establish the flux upper limits.

\textit{Systematics and Statistical Framework.}---%
Systematic uncertainties are evaluated separately for the data-driven background and the simulated signal components. 
For the background, the normalization is treated as a free-floating parameter in the global fit, leaving shape variations 
as the primary systematic source. For the signal, the hadronic model uncertainties encompass both the overall selection efficiency 
and the differential distribution shapes.

The background shape uncertainty is evaluated by constructing a symmetric systematic envelope from the maximum bin-by-bin deviations 
among alternative control templates extracted across adjacent $R_{\mu e}$ cut slices 
in CR below $-2.6$ with a step of $0.05$  (see detailed description in End Matter and Fig.~\ref{fig:shape_stability}). 
A Gaussian smoothing algorithm is subsequently applied to mitigate local statistical fluctuations, defining the continuous shape variations.

Signal simulation uncertainties primarily originate from the choice of the hadronic interaction model, 
which dictates the simulated shower development. To quantify this model dependence, 
the selection efficiencies of the combined $E_{\mathrm{rec}}$ and $R_{\mu e}$ cuts are cross-evaluated using alternative generators, 
including EPOS-LHC v3400~\cite{Pierog:2013ria} and SIBYLL 2.3d~\cite{Riehn:2019jet}. 
The maximum spread among these hadronic models is assigned as the normalization uncertainty on the signal efficiency,
yielding $4.1\%$, $9.3\%$, and $1.8\%$ for $\gamma = 10^5$, $10^6$, and $10^7$, respectively. 
In addition, due to the limited zenith-angle range in the simulation, a conservative $2\%$ uncertainty
is assigned to the signal efficiency to account for possible event migration across the selection boundary.
Similarly, systematic shape variations in the signal MLP output distribution are parameterized using the maximum bin-by-bin 
discrepancy between these alternative models. 

The statistical inference is performed using a binned profile-likelihood fit 
to the MLP score distribution. The likelihood function is defined as

\begin{equation}
\begin{split}
\mathcal{L}(\mu,\mu_{\mathrm{bkg}},
            \boldsymbol{\eta},\boldsymbol{\theta})
= {} &
\prod_{i=1}^{N_{\mathrm{bins}}}
\mathrm{Pois}(n_i \mid \nu_i)
\times
\prod_{k}
\mathcal{LN}(\eta_k \mid 1,\sigma_k)
\\
& \times
\prod_{m\in\{\mathrm{sig},\mathrm{bkg}\}}
\prod_{j}
\mathcal{G}(\theta_{m,j} \mid 0,1),
\end{split}
\end{equation}
where $n_i$ is the observed event count in the $i$-th bin, 
and the expected yield is 
$\nu_i=\mu\,s_i(\boldsymbol{\eta},\boldsymbol{\theta}_{\mathrm{sig}})+\mu_{\mathrm{bkg}}\,b_i(\boldsymbol{\theta}_{\mathrm{bkg}})$. 
Here, $\mu$ denotes the signal strength, $\mu_{\mathrm{bkg}}$ is an unconstrained free-floating multiplier for the background normalization, 
$\eta_k$ denotes the nuisance parameter associated with the $k$-th multiplicative efficiency uncertainty, 
and $\theta_{m,j}$ denotes the $j$-th template-shape nuisance parameter for $m\in\{\mathrm{sig},\mathrm{bkg}\}$.
The multiplicative efficiency uncertainties are constrained by the log-normal terms $\mathcal{LN}$,
while the template-shape uncertainties are constrained by the Gaussian terms $\mathcal{G}$ and implemented through template morphing.
For an isotropic reference flux $\Phi_{\rm ref}$, the signal template $s_i$ is normalized using the zenith-dependent exposure 
over 1438.81 live days, the projected MC generation area, and the simulated trigger, reconstruction, and selection efficiencies. 
The 90\% C.L. flux limit is then given by $\Phi_{90}=\mu_{90}\Phi_{\rm ref}$. The background normalization factor $\mu_{\mathrm{bkg}}$ 
is introduced to compensate for residual yield biases predicted by the data-driven extrapolation, and its post-fit values remain close to unity in the global fit. 

To derive the upper limit on the signal strength, 
we employ the profile-likelihood-ratio test statistic 
$q_\mu=-2\ln\!\left[\mathcal{L}\!\left(\mu,\hat{\hat{\boldsymbol{\psi}}}(\mu)\right)/\mathcal{L}\!\left(\hat{\mu},\hat{\boldsymbol{\psi}}\right)\right]$, 
where $\boldsymbol{\psi}\equiv(\mu_{\mathrm{bkg}},\boldsymbol{\eta},\boldsymbol{\theta})$ collectively 
denotes all parameters profiled for a given value of $\mu$. The quantity $\hat{\hat{\boldsymbol{\psi}}}(\mu)$ denotes 
the conditional maximum-likelihood estimate of these parameters for fixed $\mu$, while $(\hat{\mu},\hat{\boldsymbol{\psi}})$ denotes 
their joint unconditional maximum-likelihood estimate~\cite{Cowan:2010js}.

\begin{figure*}[htbp]
    \centering
    \begin{subfigure}[b]{0.32\textwidth}
        \centering
        \includegraphics[width=\textwidth]{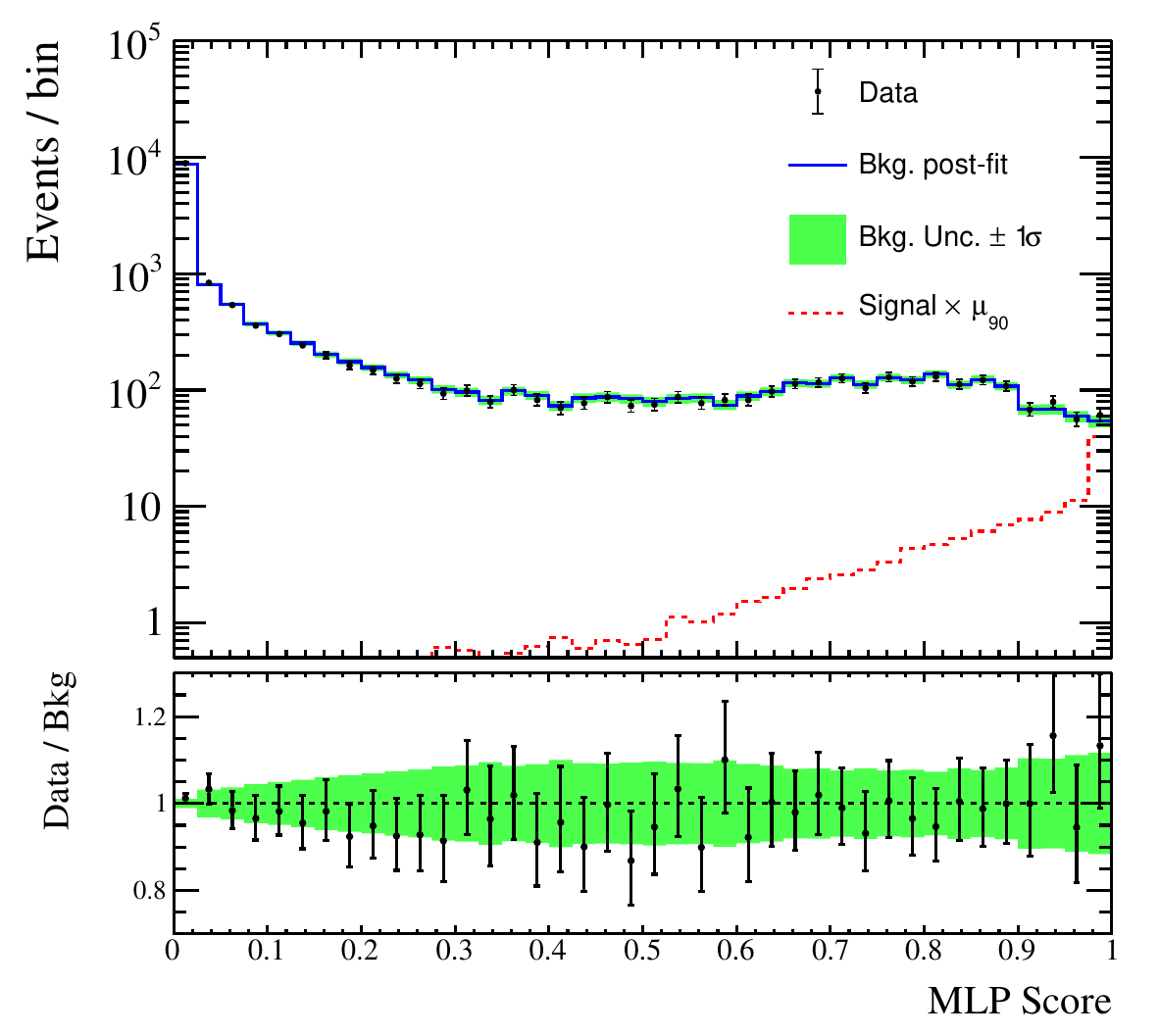}
        \caption{$\log_{10}(\gamma) = 5$}
        \label{fig:post_fit_g5}
    \end{subfigure}
    \hfill
    \begin{subfigure}[b]{0.32\textwidth}
        \centering
        \includegraphics[width=\textwidth]{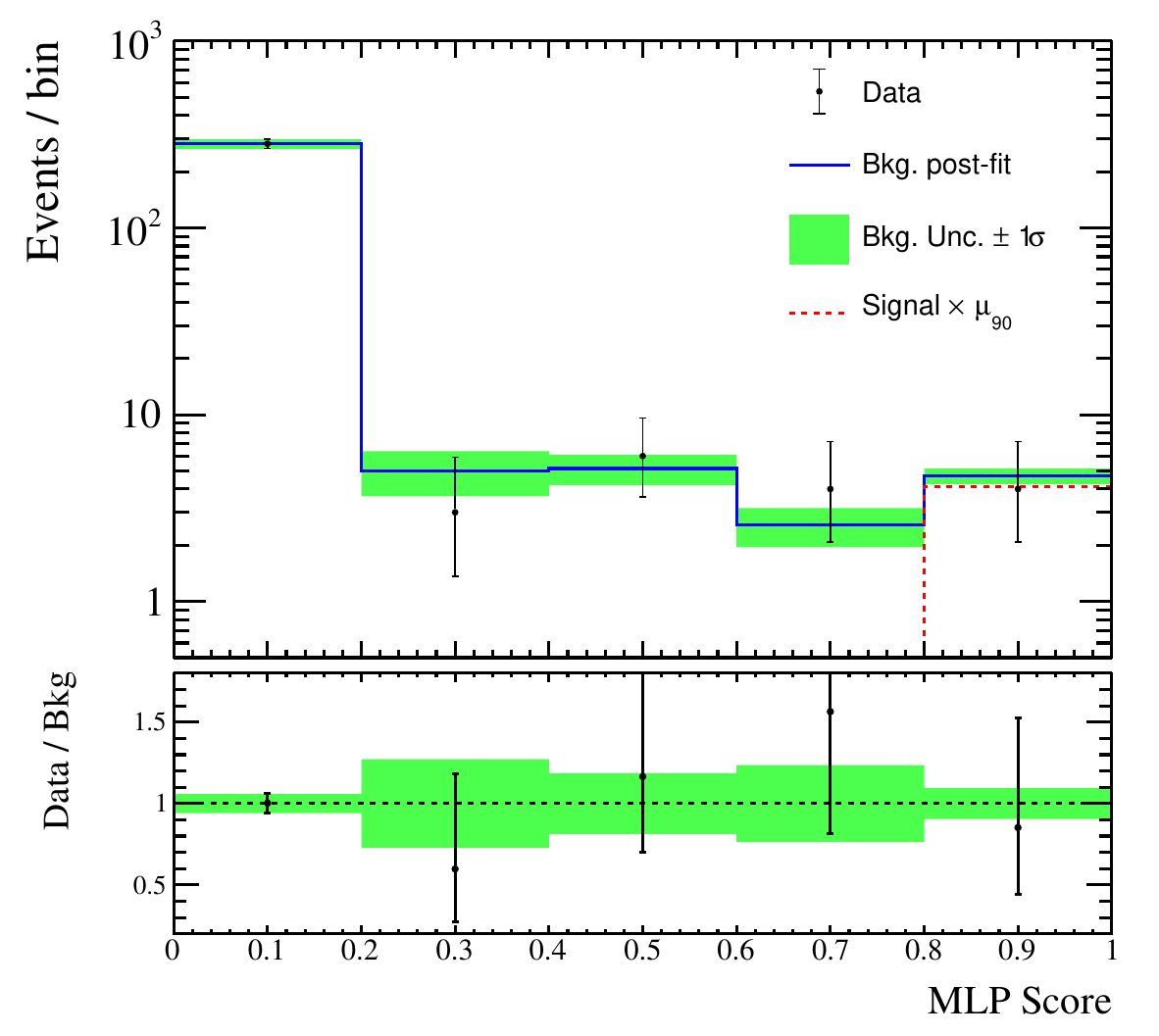}
        \caption{$\log_{10}(\gamma) = 6$}
        \label{fig:post_fit_g6}
    \end{subfigure}
    \hfill
    \begin{subfigure}[b]{0.32\textwidth}
        \centering
        \includegraphics[width=\textwidth]{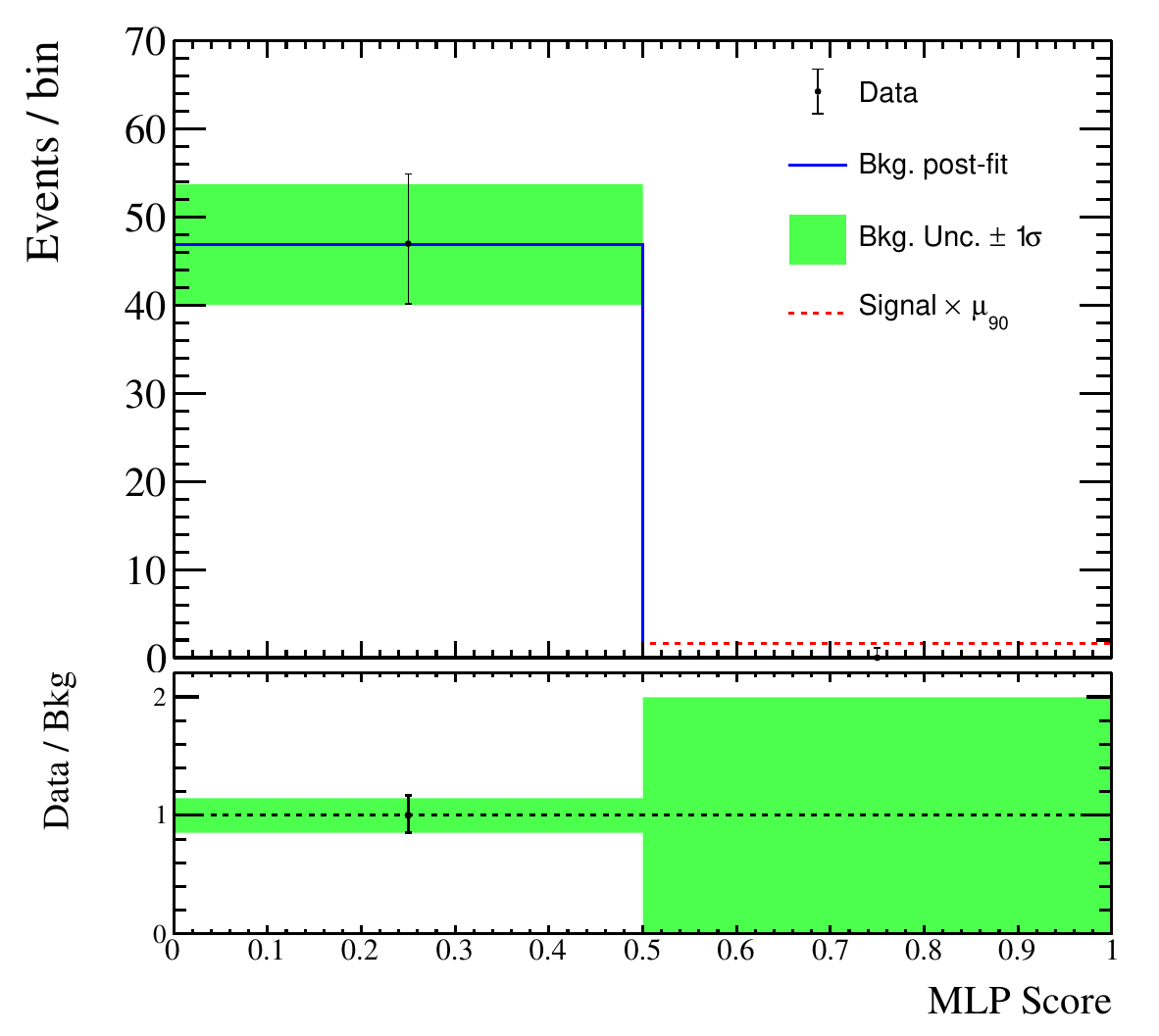}
        \caption{$\log_{10}(\gamma) = 7$}
        \label{fig:post_fit_g7}
    \end{subfigure}
    \caption{\label{fig:post_fit}Post-fit MLP score distributions for $\log_{10}(\gamma) = $ (a) 5, (b) 6, and (c) 7. Data points are compared with background-only fit expectation histograms; shaded bands indicate the total $\pm1\sigma$ uncertainties.}
\end{figure*}

\textit{Results.}---%
No significant excess over the background expectation is observed in the MLP score distributions. 
To assess the compatibility of the data with the background-only hypothesis, the post-fit distributions 
for discrete Lorentz factors $\log_{10}(\gamma)=5$, $6$, and $7$ are presented in Figs.~\ref{fig:post_fit}(a)--\ref{fig:post_fit}(c). 
The experimental data show good agreement with the background predictions within the total post-fit $\pm1\sigma$ uncertainty bands.

\begin{table}[htbp]
    \caption{\label{tab:final_limits}Expected ($\pm 1\sigma$) and observed $90\%$ C.L. upper limits on the isotropic magnetic monopole flux.}
    \begin{ruledtabular}
    \begin{tabular}{ccc}
    $\log_{10}(\gamma)$ & Expected Limit & Observed Limit \\
     & \multicolumn{2}{c}{[$10^{-19}~\text{cm}^{-2}\,\text{s}^{-1}\,\text{sr}^{-1}$]} \\
    \hline\\[-1.5ex]
    $5$ & $\LimitFiveExp$  & \LimitFiveObs  \\[1.2ex] 
    $6$ & $\LimitSixExp$   & \LimitSixObs   \\[1.2ex]
    $7$ & $\LimitSevenExp$  & \LimitSevenObs  \\
    \end{tabular}
    \end{ruledtabular}
\end{table}

\begin{figure}[htbp]
    \centering
    \includegraphics[width=0.48\textwidth]{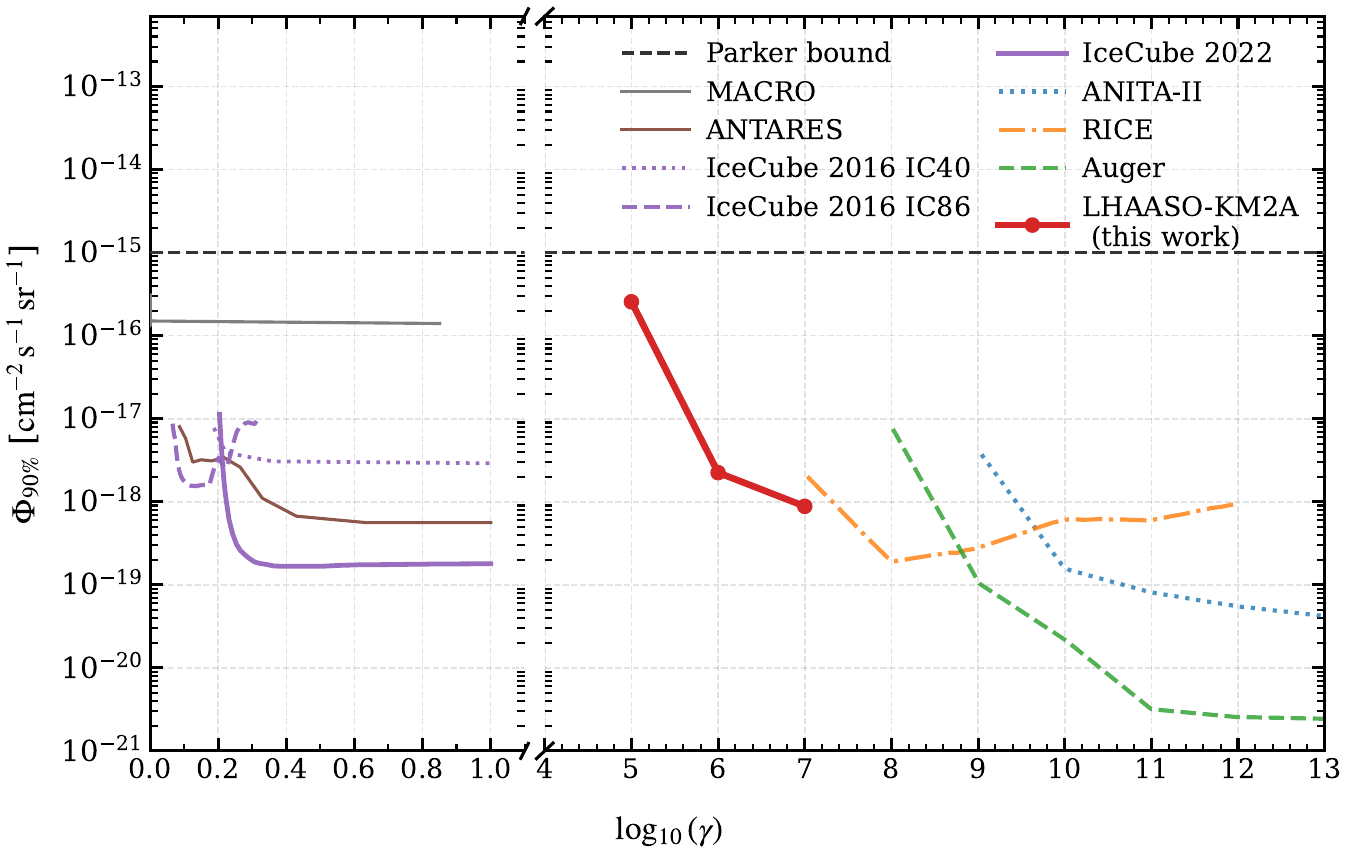}
    \caption{\label{fig:monopole_flux_limits}The $90\%$ C.L. upper limits on the magnetic monopole flux from LHAASO-KM2A 
    (red solid line), compared with the astrophysical Parker bound~\cite{Parker1970} and 
    existing constraints from other experiments.}
\end{figure}

Consequently, upper limits on the isotropic magnetic monopole flux are established at the $90\%$ confidence level (C.L.) 
using the $\text{CL}_s$ method~\cite{Read2002}. 
The expected ($\pm1\sigma$) and observed limits for a reference mass of $M = 10^{5}$~GeV are summarized in 
Table~\ref{tab:final_limits}. Figure~\ref{fig:monopole_flux_limits} presents the LHAASO-KM2A upper limits in comparison 
with the astrophysical Parker bound~\cite{Parker1970} and existing constraints from 
Pierre Auger~\cite{Auger2016}, RICE~\cite{RICE2006}, ANITA-II~\cite{ANITA2011}, MACRO~\cite{MACRO2002}, IceCube~\cite{IceCube2016,IceCube2022} 
and ANTARES~\cite{ANTARES2017}.

\textit{Conclusion.}---
We have performed the first search for ultra-relativistic MMs using the fully deployed LHAASO-KM2A array. 
Focusing on a reference monopole mass of $M = 10^{5}$~GeV, this analysis probes a critical 
kinematic window spanning Lorentz factors $10^5 \le \gamma \le 10^7$, which remains valid for other mass scales 
where pair production and photonuclear interactions dominate and atmospheric deceleration is negligible. By leveraging a strict secondary muon number selection, 
a data-driven background estimation and an MLP multivariate discriminant, the overwhelming hadronic cosmic-ray background 
is strongly suppressed. For $\gamma=10^7$, no events are observed in the high-MLP-score region.
Upon unblinding, no statistically significant signal excess is observed compared with the background-only hypothesis. 

The resulting $90\%$ C.L. upper limits set a stringent experimental boundary on the isotropic cosmic monopole flux.
At the highest boost factor investigated ($\gamma = 10^7$), the derived limit of \fluxseven{} directly constrains 
models of astrophysical monopole production, matching and extending the experimental reach of ultra-high-energy 
cosmic-ray observatories such as Pierre Auger and RICE. At lower Lorentz factors ($\gamma = 10^5$, $10^6$)
where radio and giant surface arrays lose detection efficiency, 
the LHAASO-KM2A constraints successfully fill a long-standing phase-space gap. Overall, 
the derived limits lie below the Parker bound at all three investigated Lorentz factors,
with an improvement of up to approximately three orders of magnitude.

\begin{acknowledgments}
We would like to thank the staff members of National Facility for Large High Altitude Air Shower Observatory (LHAASO) (https://cstr.cn/31117.02.LHAASO) 
who work at the LHAASO site over 4400 meter above the sea level year round to maintain the detector and keep the water recycling system, 
electricity power supply and other components of the experiment operating smoothly.
We are grateful to Chengdu Management Committee of Tianfu New Area for the constant financial support 
for research with LHAASO data. 
We appreciate the computing and data service support provided by the National High Energy Physics Data Center 
for the data analysis in this paper. 
This research work is supported by the following grants: 
The National Natural Science Foundation of China: 
No.12588101, No.12393854, 
No.12220101003, No.12175121, No.12275280, No.12393851, No.12393852, 
No.12393853,  No.12205314, No.12105301, No.12305120, No.12261160362, No.12105294, No.U1931201, No.12375107, 
and the CAS Project for Young Scientists in Basic Research (No.YSBR-061) and 
from the NSRF via the Research and Innovation Acceleration Agency for Competitiveness and Area Development (RCAD) 
(Program Management Unit for Technology and Innovation for Future Industries (PMU-B) : Brainpower for Future Industries) [grant number B39G690003].
\end{acknowledgments}

%

\clearpage
\appendix

\begin{figure}[htbp]
    \centering
    \includegraphics[width=0.45\textwidth]{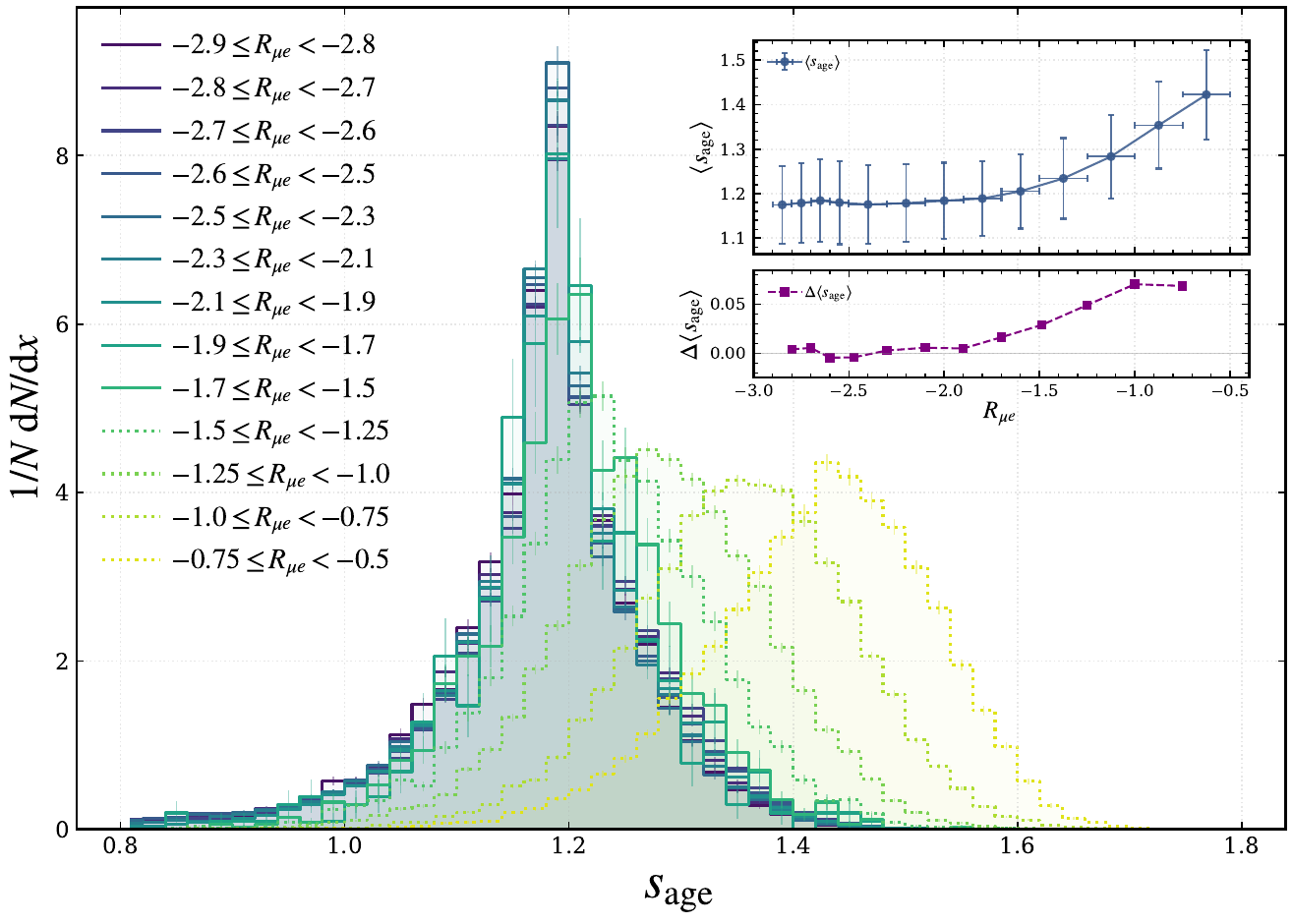}
    \caption{\label{fig:asymptotic_scan}Normalized shower age ($s_{\text{age}}$) distribution scanned across different $R_{\mu e}$ slices.}
\end{figure}

\textit{Background Modeling Validation for $\gamma=10^5$.}---%
The data-driven background estimation assumes shape invariance of the multi-variable shower parameters across 
adjacent $R_{\mu e}$ sidebands. Figure~\ref{fig:asymptotic_scan} illustrates this behavior using 
the normalized shower-age distributions in successive $R_{\mu e}$ slices, where the evolution of the distribution is most apparent.

The core physical mechanism governing this convergence is rooted in the high-energy fluctuations of cosmic-ray air showers. 
As mapped by the $s_{\text{age}}$ distribution and its variation in Fig.~\ref{fig:asymptotic_scan}, 
the longitudinal shower development exhibits large structural mutations in the high-muon regimes ($R_{\mu e} \ge -1.5$, yellow dotted lines). 
However, as $R_{\mu e}$ continuously decreases, the secondary hadronic interaction is depleted, 
forcing the atmospheric cascades into the pure electromagnetic fluctuation limit. 
Below the critical boundary of $R_{\mu e} \le -2.6$, the longitudinal profiles asymptotically freeze into 
an invariant configuration (dark solid lines). This stabilization is formalized in the inset panels: 
the mean profile $\langle s_{\text{age}} \rangle$ flattens into a rigid plateau, with its nominal residual variation 
$\Delta \langle s_{\text{age}} \rangle$ bounded within $0.01$. 
We define these $R_{\mu e}$ slices as alternative adjacent templates included in the systematics evaluation.

This shape stability inside the deep control region also holds for the other five observables shown in Fig.~\ref{fig:shape_stability}. 
For the other two phase points $\gamma=10^6, 10^7$, observable distributions are shown in Supplemental Material. 
By capitalizing on this asymptotic convergence, 
the empirical template extracted from the sideband is used to estimate the background shape in the signal region. 
The small variations across adjacent sidebands are conservatively treated as a shape systematic uncertainty in the profile likelihood fit.
In contrast, the simulated monopole signal shows a completely different distribution across all variables, 
confirming the strong discrimination power.

\textit{MLP Inputs Definition.}---%
To separate the hypothetical monopole signal from cosmic-ray background, an MLP classifier is trained using six shower observables. 
Let $q_i$ denote the total integrated charge (in photoelectrons, PE) registered by the $i$-th active ED unit, 
and $r_i = \sqrt{(x_i - x_{\mathrm{core}})^2 + (y_i - y_{\mathrm{core}})^2}$ represent its lateral distance to the reconstructed shower core. 
The definitions are as follows:

\begin{figure*}[htbp]
    \centering
    \includegraphics[width=0.95\textwidth]{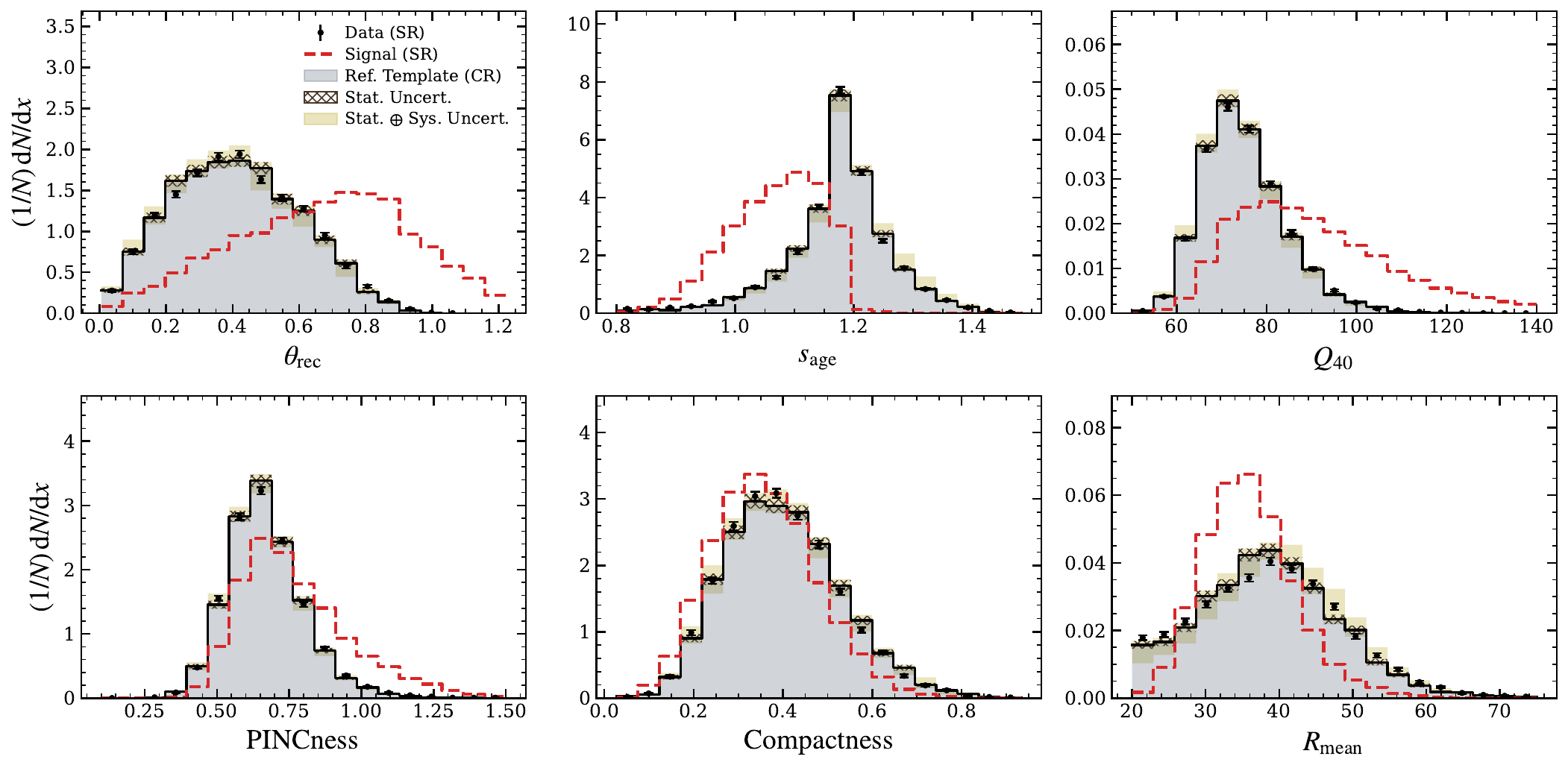}
    \caption{\label{fig:shape_stability}Validation of the shower shape invariance across the six MLP input observables for the $\gamma = 10^5$ analysis. The reference template (black step) is compared against adjacent control slices (incorporated in yellow shadow) and the simulated monopole signal topology (red dashed line). The data in SR (black dots) show excellent consistency with templates extracted from CR. }
\end{figure*}

\noindent\textbf{(1) Reconstructed Zenith Angle ($\theta_{\mathrm{rec}}$):} Reconstructed via a timing fit to the relative arrival times of secondary particles across the active array. Large zenith angles provide a deep atmospheric filter that suppresses standard hadronic backgrounds. 

\noindent\textbf{(2) Shower Age ($s_{\mathrm{age}}$):} A shape parameter that indicates the steepness of the lateral distribution function, extracted by fitting the recorded lateral particle densities with a modified NKG function~\cite{Kamata1958,Greisen:1960progress}. 

\noindent\textbf{(3) Outer Core Density ($Q_{40}$):} The average charge per hit registered in the outer core region beyond a radial boundary of $40\,\text{m}$, defined as $Q_{40} = \sum_{i \in \mathbb{I}_{40}} q_i / N_{40}$, where $\mathbb{I}_{40} = \{i \mid r_i > 40\,\text{m}\}$ and $N_{40} = |\mathbb{I}_{40}|$.

\noindent\textbf{(4) PINCness:} The shower is segmented into concentric radial bins $b$ with a width of $10\,\text{m}$, 
adapted from reference~\cite{Abeysekara_2017}. $\text{PINCness} = \frac{1}{N_{\mathrm{valid}}} \sum_{b} \sum_{i \in \text{bin } b} [(\zeta_i - \bar{\zeta}_b)/\sigma_b]^2$, 
where $\zeta_i = \log_{10}(q_i + 1)$, 
and $N_{\mathrm{valid}}$ is the cumulative number of hits inside all valid bins.
$\bar{\zeta}_{b}$ is the mean $\zeta$ value in radial bin $b$ for the event under evaluation, and $\sigma_{b}$ 
is the fixed standard deviation in that bin calculated from all signal MC events at the corresponding energy point.

\noindent\textbf{(5) Compactness:} The ratio of the total hit multiplicity to the maximum localized charge recorded at a large lateral radius ($r > 45\,\text{m}$), defined as $\text{Compactness} = N_{\mathrm{hit}} / \max_{i \in \mathbb{I}_{45}} (q_i)$ where $\mathbb{I}_{45} = \{i \mid r_i > 45\,\text{m}\}$, adapted from reference~\cite{ALFARO2022}. Monopole events exhibit higher compactness values due to the absence of peripheral hadronic sub-cores.

\noindent\textbf{(6) Mean Radius ($R_{\mathrm{mean}}$):} The charge-weighted mean lateral distance of all triggered ED units, parameterizing the global lateral dispersion of the energy deposition: $R_{\mathrm{mean}} = \sum_{i=1}^{N_{\mathrm{hit}}} q_i r_i / \sum_{i=1}^{N_{\mathrm{hit}}} q_i$, where $N_{\mathrm{hit}}$ is the total number of hit detectors.

\begin{figure*}[tbp]
    \centering
    \begin{subfigure}{\textwidth}
        \centering
        \includegraphics[width=0.9\textwidth]{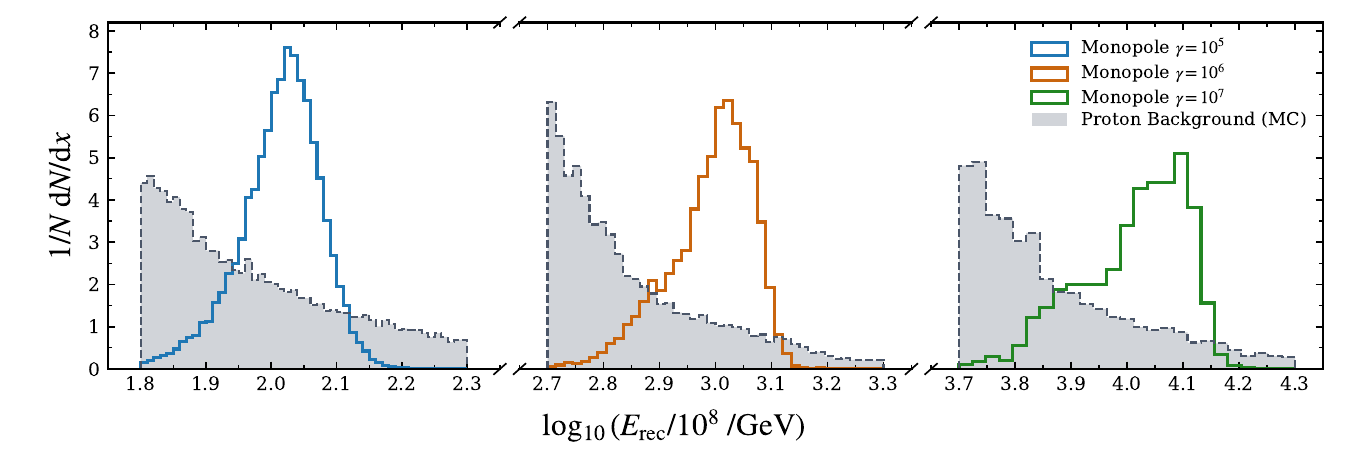}
       
        \label{fig:sr_recE}
    \end{subfigure}
    
    \begin{subfigure}{\textwidth}
        \centering
        \includegraphics[width=0.9\textwidth]{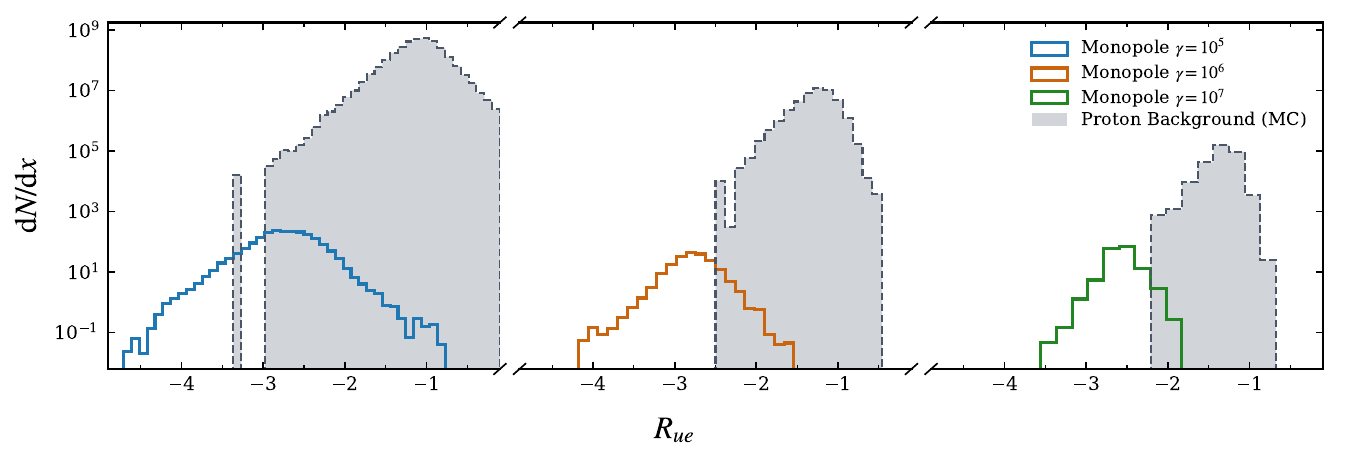}
       
        \label{fig:sr_rue}
    \end{subfigure}
    \caption{\label{fig:sr_variables}Distributions of the primary selection variables for the simulated magnetic monopole signals at $\gamma = 10^5$, $10^6$, and $10^7$ compared against the simulated proton with Gaisser composition weight: reconstructed energy $\log_{10}(E_{\text{rec}}/10^8\,\text{GeV})$ (top) and muon-to-electromagnetic ratio $R_{\mu e}$ (bottom).}
\end{figure*}

\end{document}


\title{Supplemental Material: Search for Ultra-Relativistic Magnetic Monopoles with LHAASO-KM2A}


\collaboration{LHAASO Collaboration}
\noaffiliation


\maketitle

\section*{Background Template Validation for High Lorentz Phase Points}
\label{sec:sup_high_gamma_validation}

\begin{figure*}[htbp]
    \centering
    \includegraphics[width=0.75\textwidth]{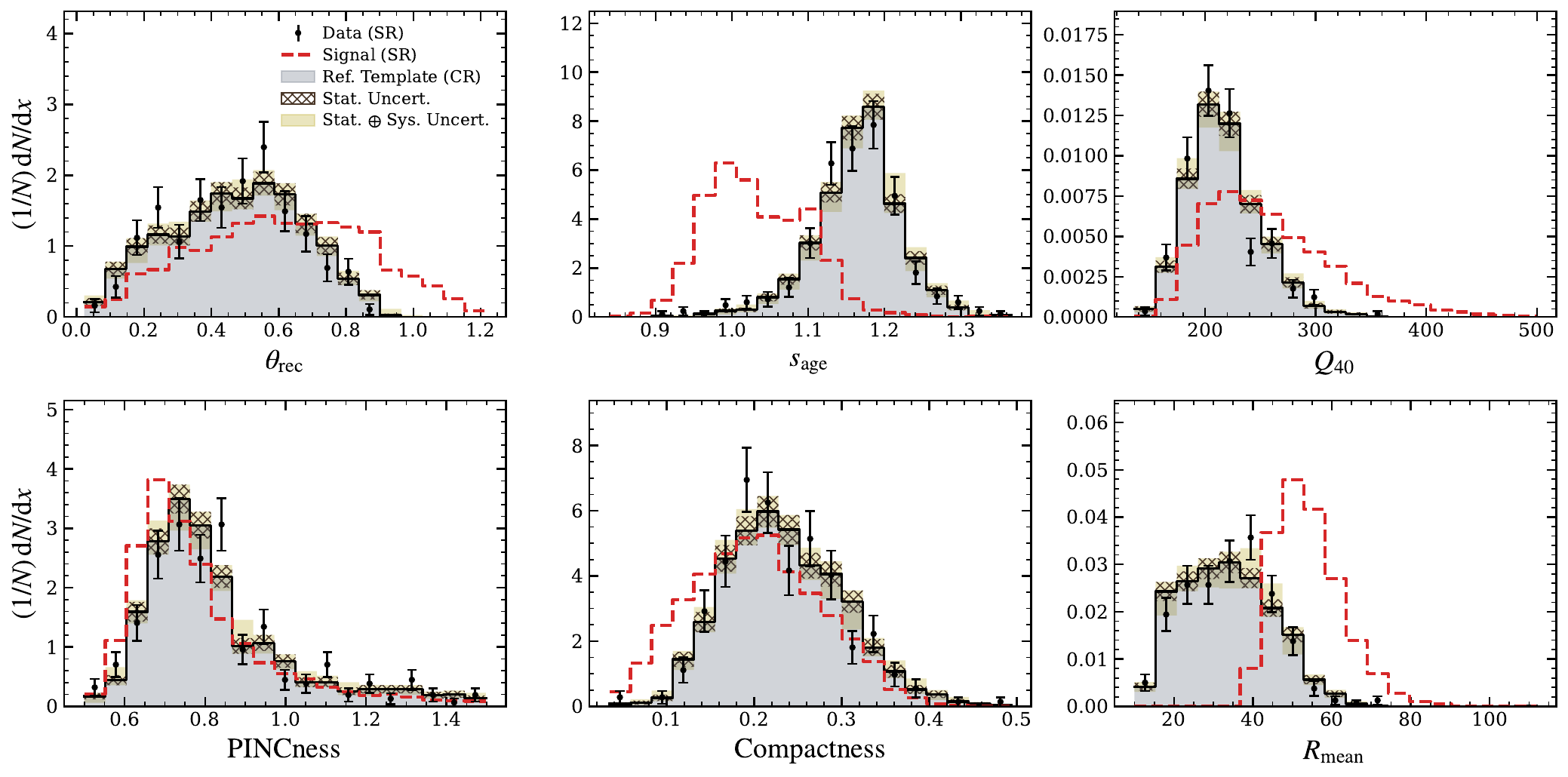}
    \caption{\label{fig:shape_stability_1e11}Validation of the shower shape invariance across the six MLP input observables for the $\gamma = 10^6$ analysis. The reference template is compared against adjacent control slices (represented by the systematic bands) and the simulated monopole signal topology (red dashed line). The observed data in the search window (black dots) are consistent with the background prediction within statistical uncertainties.}
\end{figure*}

\begin{figure*}[htbp]
    \centering
    \includegraphics[width=0.75\textwidth]{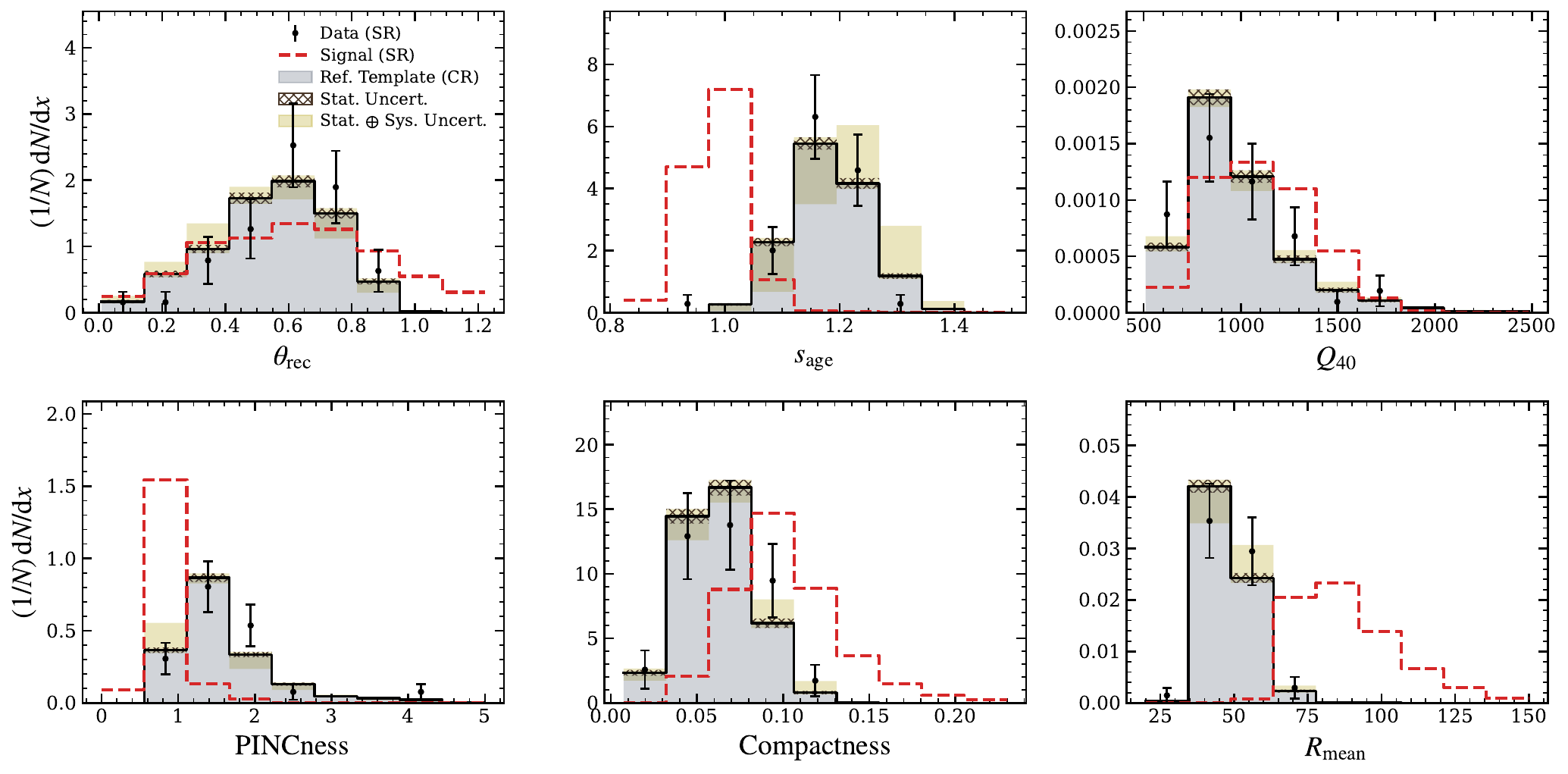}
    \caption{\label{fig:shape_stability_1e12}Validation of the shower shape invariance across the six MLP input observables for the $\gamma = 10^7$ analysis.}
\end{figure*}

The invariance of the data-driven background templates is evaluated for the higher energy regimes corresponding to $\gamma = 10^6$ and $10^7$. Figures~\ref{fig:shape_stability_1e11} and \ref{fig:shape_stability_1e12} present the multi-variable validation across the six Multi-Layer Perceptron (MLP) input observables for these two phase points, respectively.

Similar to the Lorentz phase point ($\gamma = 10^5$), the reference background template derived from the deep control region (CR) successfully conform to the shapes of the adjacent control slices. The systematic variations across these sliding sidebands are bounded by the predefined symmetric shape systematic envelope.

